\documentclass[fleqn,twoside]{article}
\usepackage{espcrc2}

\usepackage{graphicx}
\usepackage[figuresright]{rotating}

\newcommand{\SQS}       {\ensuremath{\sqrt{s_{\rm ee}}}}
\newcommand{\GV}        {\ensuremath{\mathrm{GeV}}}
\newcommand{\EE}        {\ensuremath{\mathrm{e}^+\mathrm{e}^-}}
\newcommand{\GG}        {\ensuremath{\mathrm{\gamma\gamma}}}
\newcommand{\PP}        {\ensuremath{\mathrm{p\bar{p}}}} 
\newcommand{\pb}        {\ensuremath{\mathrm{pb}}}

\newcommand{\PT}        {\ensuremath{p_{\perp}}}
\newcommand{\dedx}      {\ensuremath{{\rm d}E/{\rm d}x}}

\newcommand{\lumi}      {\ensuremath{\mathcal{L}}}
\newcommand{\costs}     {\ensuremath{\mathrm{|\cos\theta^{*}|}}}
\newcommand{\cost}      {\ensuremath{\mathrm{|\cos\theta|}}}
\newcommand{\NE}        {\ensuremath{N_{\rm ev}}}
\newcommand{\EDET}      {\ensuremath{\mathrm{\varepsilon}_{\rm DET}}}
\newcommand{\ETRIG}     {\ensuremath{\mathrm{\varepsilon}_{\rm TRIG}}}
\newcommand{\spts}      {\ensuremath{|\sum{\vec{p}_\perp}|^{2}}}
\newcommand{\SI}        {\ensuremath{\mathrm{\sigma}}}

\newcommand{\AmS}{{\protect\the\textfont2
  A\kern-.1667em\lower.5ex\hbox{M}\kern-.125emS}}

\title{Measurement of the Cross-Section for the Process $\GG\to\PP$ at $\SQS = 183-189\,\GV$ with the OPAL Detector at LEP}

\author{T. Barillari\address[MPI]{Max-Planck-Institut f\"{u}r Physik, 
        Werner-Heisenberg-Institut, \\ 
         Foehringer Ring 6, D-80805 Muenchen, Germany}}
\begin{document}

\begin{abstract}
 The exclusive production of proton-antiproton pairs in the collisions
 of two quasi-real photons has been studied using 
 data taken at $\SQS = 183\,\GV$ and $189\,\GV$ with the OPAL detector at 
 LEP.
 Results are presented for $\PP$ invariant masses, $W$, in the range 
 $2.15<W<3.95\,\GV$. 
 The cross-section measurements are compared with previous data and with 
 recent analytic calculations based on the quark-diquark model.
\vspace{1pc}
\end{abstract}

\maketitle

\section{INTRODUCTION}
The exclusive production of proton-antiproton (\PP) pairs in the collision of
two quasi-real photons can be used to test predictions of
QCD. At LEP the photons are emitted by the beam electrons\footnote{In
this paper positrons are also referred to as electrons.}
and the $\PP$ pairs are produced in the process $\EE\to\EE\gamma\gamma\to\EE\PP$.

The application of QCD to exclusive photon-photon reactions
is based on the work of Brodsky and Lepage~\cite{Lepage:1980fj}. 
Calculations based on this ansatz~\cite{Farrar:1985gv,Millers:1986ca}
use a specific model of the proton's three-quark wave function by 
Chernyak and Zhitnitsky~\cite{Chernyak:1984bm}.
This calculation yields cross-sections about one order of magnitude smaller 
than the existing experimental 
results~\cite{Althoff:1983pf,Bartel:1986sy,Aihara:1987ha,Albrecht:1989hz,Artuso:1994xk,Hamasaki:1997cy,OPALpap}, 
for $\PP$ centre-of-mass energies $W$ greater than $2.5\,\GV$.

To model non-perturbative effects, the introduction
of quark-diquark systems has been proposed~\cite{Ansel:1987vk}. 

Recent studies~\cite{berger:1997} have extended the systematic 
investigation of hard exclusive reactions within the quark-diquark model 
to photon-photon processes~\cite{Anselmino:1989gu,Kroll:1991a,Kroll:1993zx,Kroll:1996pv}.

The calculations of the integrated cross-section for the process $\GG\to\PP$
in the angular range $\costs < 0.6$ (where $\theta^{*}$ is the angle between 
the proton's momentum and the electron beam direction in the  
$\PP$ centre-of-mass system) and for $W > 2.5\,\GV$ are 
in good agreement with experimental 
results~\cite{Artuso:1994xk,Hamasaki:1997cy}, whereas the pure quark model 
predicts much smaller cross-sections~\cite{Farrar:1985gv,Millers:1986ca}.  

In this paper,  we present a measurement of the cross-section~\cite{OPALpap} 
for the exclusive process $\EE\to\EE \PP$ in the range $2.15<W<3.95\,\GV$, using 
data taken with the OPAL detector~\cite{Ahmet:1991eg} 
at $\SQS = 183\,\GV$ and $189\,\GV$ at LEP. 
The integrated luminosities for the two energies are $62.8\,\pb^{-1}$ and 
$186.2\,\pb^{-1}$.

\section{EVENT SELECTION }

The $\EE\to\EE\PP$ events are selected by the following set of cuts:
\begin{enumerate}
\itemsep -1pt
\item
     The sum of the energies measured in the barrel and endcap sections
     of the electromagnetic calorimeter must be less than half the beam
     energy.  
\itemsep -1pt
\item
     Exactly two oppositely charged tracks are required with each track
     having at least 20 hits in the central jet chamber to ensure
     a reliable determination of the specific energy loss ${\rm d}E/{\rm d}x$.
     The point of closest approach to the interaction point
     must be less than 1~cm in the $r\phi$ plane 
     and less than 50~cm in the $z$ direction.
\itemsep -1pt
\item
     For each track the polar angle must be in the range $\cost < 0.75$ 
     and the transverse momentum
     $\PT$ must be larger than $0.4\,\GV$.
     These cuts ensure a high 
     trigger efficiency and good particle identification.
\itemsep -1pt
\item 
     The invariant mass $W$ of the $\PP$ final state
     must be in the range $2.15<W<3.95\,\GV$. 
     The invariant mass is determined from the measured 
     momenta of the two tracks using the proton mass.
\itemsep -1pt
\item
     The events are boosted into the rest system of the
     measured $\PP$ final state.
     The scattering angle of the tracks in this
     system has to satisfy $\costs < 0.6$.
\itemsep -1pt
\item 
     All events must fulfil the trigger
     conditions described in~\cite{OPALpap}.
\itemsep -1pt
\item
     The large background from other exclusive processes, 
     mainly the production of e$^+$e$^-$, $\mu^+\mu^-$, and $\pi^+\pi^-$
     pairs, is reduced by particle identification 
     using the specific energy loss $\dedx$ in the jet
     chamber and the energy in the electromagnetic calorimeter. The 
     $\dedx$ probabilities of the tracks must be consistent with the 
     p and $\overline{\mbox{p}}$ hypothesis. 
     \begin{itemize}
     \item[-]
      Events where the ratio $E/p$ for each track lies in the range 
      $0.4 < E/p < 1.8$ are regarded as possible $\EE\to\EE\EE$ candidates, 
      $E$ here is the energy of the ECAL cluster 
      associated with the track with momentum~$p$.
      These events are rejected if the 
      $\dedx$ probabilities of the two tracks are consistent 
      with the electron hypothesis.
      \item[-]
      Events where the ratio $E/p$ for each track is less than $0.8$, 
      as expected for a minimum ionizing particle, are regarded as possible
      background from $\EE\to\EE\mu^+\mu^-$ events. This background is 
      reduced by rejecting events where the $\dedx$ probability for
      both tracks is consistent with the muon hypothesis. 
      This cut is also effective in reducing the $\pi^{+}\pi^{-}$
      background.      
     \item[-]
      The $\dedx$ probability for the proton hypothesis has to be greater than 
      $0.1\%$ for each track and it has to be larger than the probabilities 
      for the pion and kaon hypotheses. 
      The product of the $\dedx$ probabilities for both 
      tracks to be (anti) protons has to be larger than the product of the 
      $\dedx$ probabilities for both tracks to be electrons.
     \end{itemize}
\itemsep -1pt
\item
     Cosmic ray background is eliminated by applying a 
     muon veto~\cite{Akers:1995vh}.
\itemsep -1pt
\item
     Exclusive two-particle final states are selected by
     requiring the transverse component of the momentum 
     sum squared of the two tracks, $\spts$, to be smaller than
     $0.04\,\GV^2$. By restricting the maximum value of $Q_i^2$, 
     this cut also ensures that the interacting photons
     are quasi-real. Therefore no further cut rejecting events with
     scattered electrons in the detector needs to be applied.
\end{enumerate}
After all cuts $163$ data events are selected, 35 events 
at $\SQS=183$~$\GV$ and $128$ events at $\SQS=189$~$\GV$.
Background from events containing particles other than (anti-)protons
is negligible due to the good rejection power of the $\dedx$ cuts.
No event remains after applying the event selection to the 
background Monte Carlo samples~\cite{OPALpap}. 
Since the $\PP$ final state is fully reconstructed, the
experimental resolution for $W$ (determined with Monte Carlo simulation)
is better than $1\%$. The experimental resolution for $\costs$ is about 
$0.014$.

\section{CROSS-SECTION MEASUREMENTS}

The differential cross-section for the process $\EE\to\EE\PP$ is given by
%\begin{equation}
\begin{eqnarray}
  \frac{{\rm d}^2\SI(\EE\to\EE\PP)}{{\rm d}W\,{\rm d}\costs}  =  \hspace*{3.0cm} \nonumber \\
  \frac{{\NE(W,\costs)}}{{\cal{L}}_{\EE}\ETRIG\,\EDET\,(W,\costs)\,\Delta W\,\Delta\costs} 
 \label{eq:diffcross}
%\end{equation}
\end{eqnarray}
where \NE\ is the number of events selected in each $(W,\costs)$
bin, $\ETRIG$ is the trigger efficiency, $\EDET$ is the detection 
efficiency, $\cal{L}_{\EE}$ is the measured integrated luminosity, and 
$\Delta W$ and $\Delta\costs$ are the bin widths in $W$ and in $\costs$.

The total cross-section $\SI(\GG\to\PP)$ for a given value of 
$\SQS$ is obtained from the differential cross-section
${\rm d}\SI(\EE\to\EE\PP)/{\rm d}W$ using the luminosity
function ${\rm d}\lumi_{\GG}/{\rm d}W$~\cite{Low:1960wv}:
\begin{equation}
 \SI(\GG\to\PP)= 
 \frac{{\rm d}\SI(\EE\to\EE\PP)}{{\rm d}W}\left/ 
 \frac{{\rm d}\lumi_{\GG}}{{\rm d}W}\right. .
 \label{eq:cross}
\end{equation}
The luminosity function ${\rm d}\lumi_{\GG}/{\rm d}W$ 
is calculated by the {\sc Galuga} program~\cite{Schuler:1996gt}.
The resulting differential cross-sections for the process $\GG\to\PP$ 
in bins of $W$ and $\costs$ are then summed over 
$\costs$ to obtain the total cross-section as a function of $W$
for $\costs<0.6$.

\section{RESULT AND DISCUSSION}

\begin{figure}[htbp]\vspace*{-0.5cm}
 \centering
  \resizebox{0.312\textwidth}{!}{%
  \includegraphics{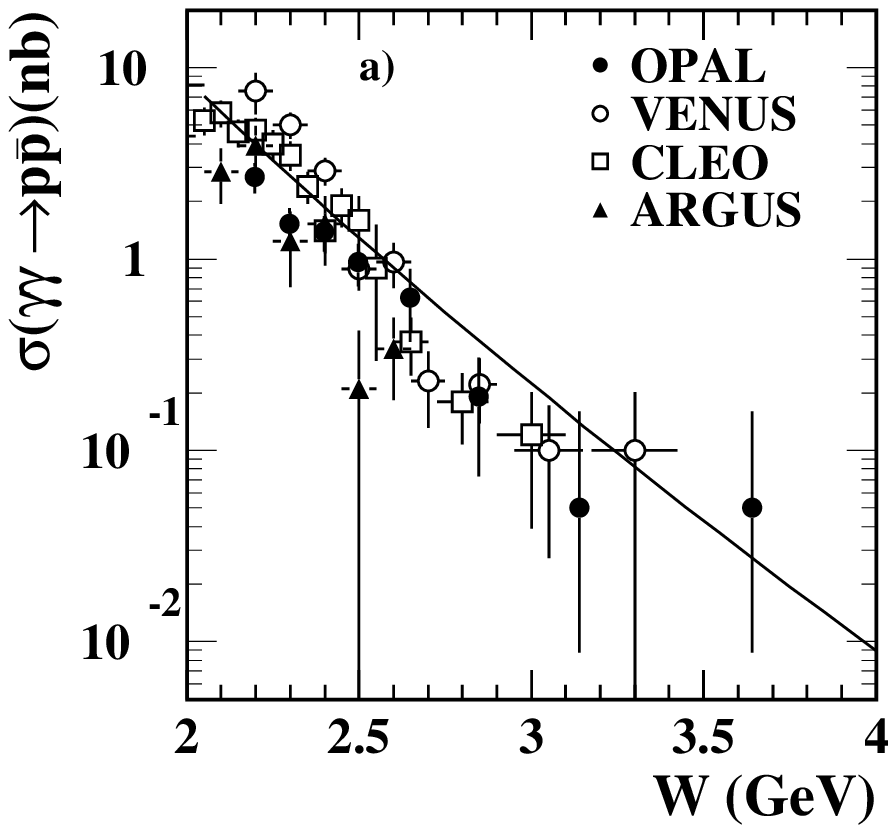}}
  \resizebox{0.312\textwidth}{!}{%
  \includegraphics{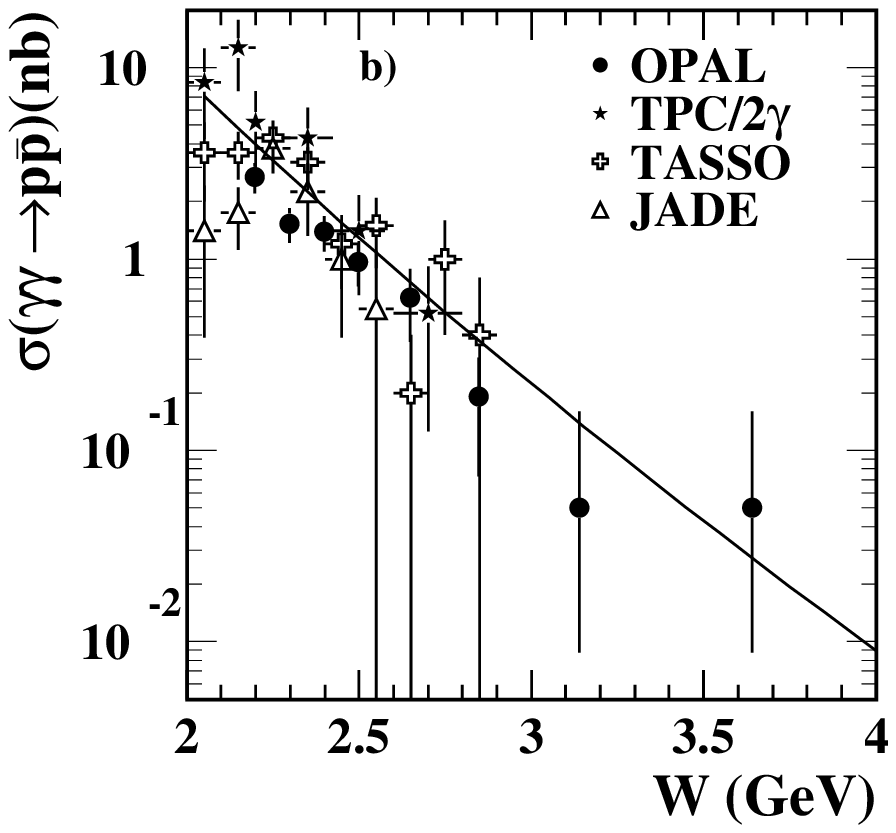}}
  \resizebox{0.312\textwidth}{!}{%
  \includegraphics{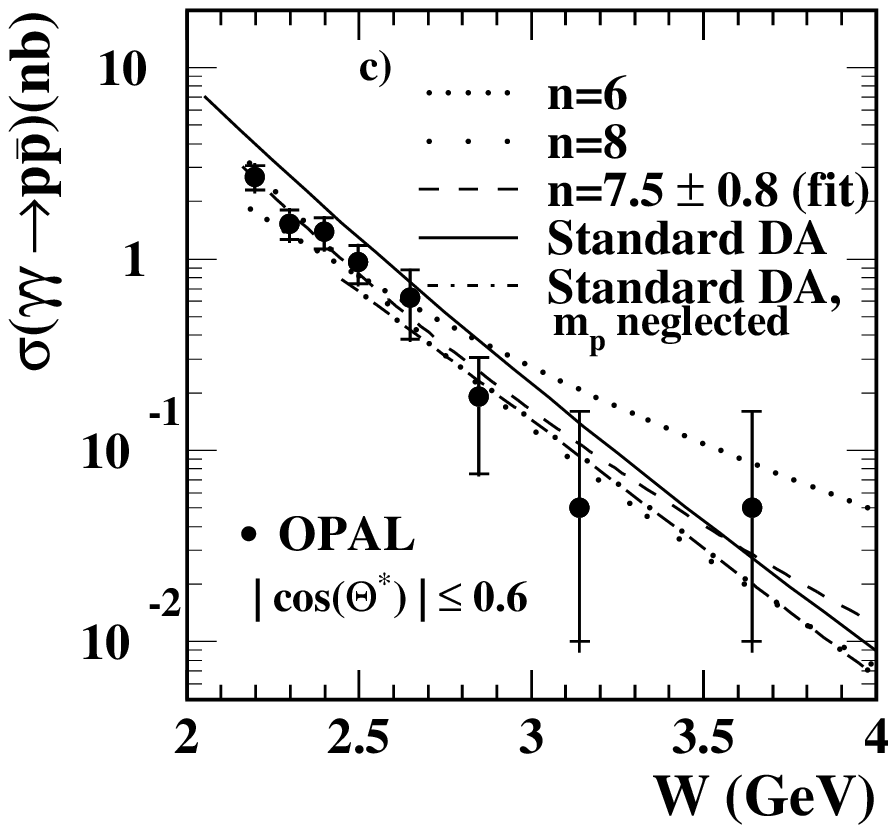}}
    \vspace*{-1.1cm}
    \caption{Cross-sections $\sigma(\GG\to\PP)$ as a function of $W$ for $\costs < 0.6$.
%        The data and the theoretical predictions cover a range of 
%        $\costs < 0.6$. 
        Data points are plotted at the value of $\langle W\rangle$.
        a,b) The data are compared to other experimental 
        results~\cite{Aihara:1987ha,Albrecht:1989hz,Artuso:1994xk,Hamasaki:1997cy}
        and to the quark-diquark model prediction~\cite{berger:1997}.
        The error bars include statistical and systematic uncertainties, 
        except for TASSO~\cite{Althoff:1983pf} where the uncertainties are
        statistical only.
	c) The data are compared to the quark-diquark model 
	of~\cite{Ansel:1987vk} (dash-dotted line), and 
	of~\cite{berger:1997} (solid line), using the standard 
	DA~\cite{Ansel:1987vk,berger:1997} with and without neglecting the 
        mass $m_{\rm p}$ of the proton, 
	and with the predictions of the power law with fixed 
	and with fitted exponent $n$. 
%	   The inner error bars are the statistical 
%	   uncertainties and the outer error bars are the total uncertainties.}
	Inner error bars are statistical 
	uncertainties and outer error bars are total uncertainties.}
    \label{fig:w}
\vspace*{-0.5cm}
\end{figure}

The measured cross-sections~\cite{OPALpap} as a function of $W$ are showed in 
Fig.~\ref{fig:w}.
The average $\langle W\rangle$ in each bin has been
determined by applying the procedure described in~\cite{Lafferty:1995}. 
The measured cross-sections $\SI(\GG\to\PP)$ 
for $2.15<W<3.95\,\GV$ and for $\costs < 0.6$ 
are compared with the results obtained by 
ARGUS~\cite{Albrecht:1989hz}, CLEO~\cite{Artuso:1994xk} and 
VENUS~\cite{Hamasaki:1997cy} in Fig.~\ref{fig:w}a
and to the results obtained by
TASSO~\cite{Althoff:1983pf}, 
JADE~\cite{Bartel:1986sy} and TPC/$2\gamma$~\cite{Aihara:1987ha}
in Fig.~\ref{fig:w}b.
The quark-diquark model predictions~\cite{berger:1997} are also shown.
Reasonable agreement is found between this measurement 
and the results obtained by other experiments for $W>2.3\,\GV$.
At lower $W$ our measurements agree with the measurements by 
JADE~\cite{Bartel:1986sy} and ARGUS~\cite{Albrecht:1989hz},
but lie below the results
obtained by CLEO~\cite{Artuso:1994xk}, and VENUS~\cite{Hamasaki:1997cy}.
The cross-section measurements reported here extend towards higher 
values of $W$ than previous results.
Fig.~\ref{fig:w}c shows the measured $\GG\to\PP$ 
cross-section as a function of $W$ together with some  
predictions based on the quark-diquark 
model~\cite{Ansel:1987vk,berger:1997}.  
There is good agreement between our results and the older
quark-diquark model predictions~\cite{Ansel:1987vk}.
The most recent calculations~\cite{berger:1997} lie above the data, 
but within the estimated theoretical uncertainties the predictions are 
in agreement with the measurement.

An important consequence of the pure quark hard scattering 
picture is the power law which follows from the dimensional counting 
rules~\cite{Brodsky:1973kr,Matveev:1973ra}.
The dimensional counting rules state that an exclusive cross-section 
at fixed angle has an energy dependence connected with the number of 
hadronic constituents participating in the process under investigation. 
We expect that for asymptotically large $W$ and fixed 
$\costs$
\begin{equation}
  \frac{{\rm d}\SI{(\GG\to\PP)}}{{\rm d}t} \sim W^{2(2-n)}
  \label{eq:powerlaw}
\end{equation}
where $n=8$ is the number of elementary fields
and $t = -W^2/2(1-\costs)$. The introduction
of diquarks modifies the power law by decreasing $n$ to $n=6$.
This power law is compared
to the data in Fig.~\ref{fig:w}c with 
$\SI(\GG\to\PP) \sim W^{-2(n-3)}$ using three 
values of the exponent $n$: fixed values $n=8$, $n=6$, 
and the fitted value 
$n=7.5\pm0.8$ obtained by taking into account statistical uncertainties only. 
More data covering a wider range of $W$ would be required to determine the 
exponent $n$ more precisely. 
\begin{figure}[htbp]\vspace*{-0.5cm}
   \centering
   \resizebox{0.35\textwidth}{!}{%
   \includegraphics{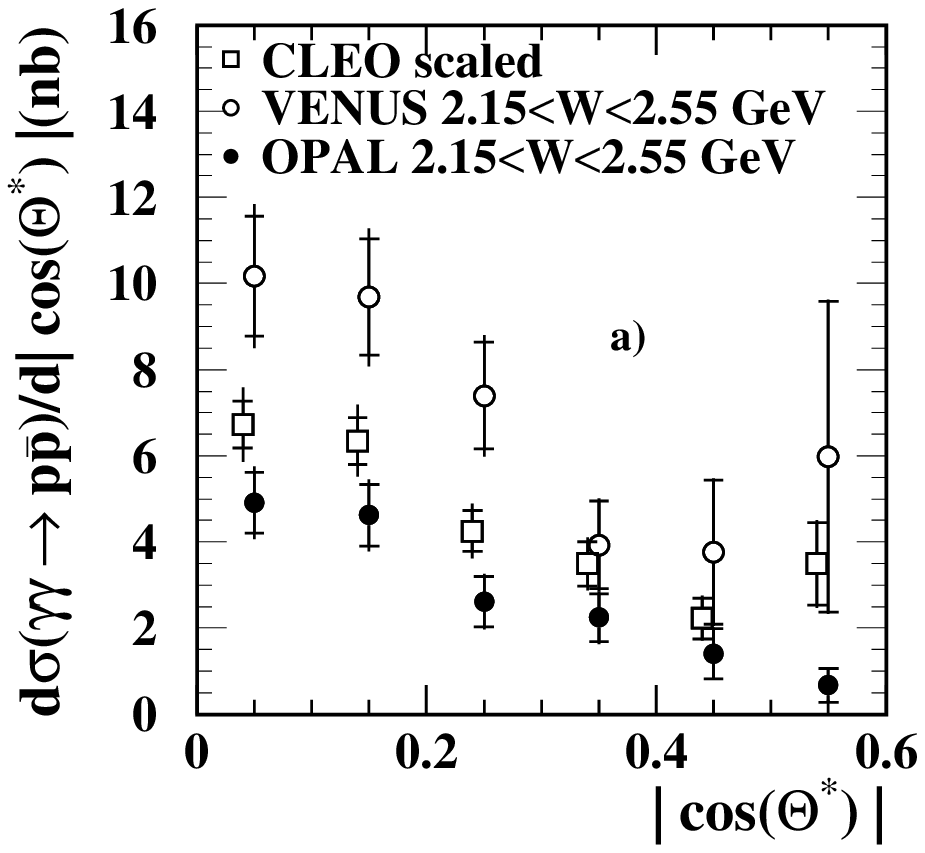}}
   \resizebox{0.35\textwidth}{!}{%
   \includegraphics{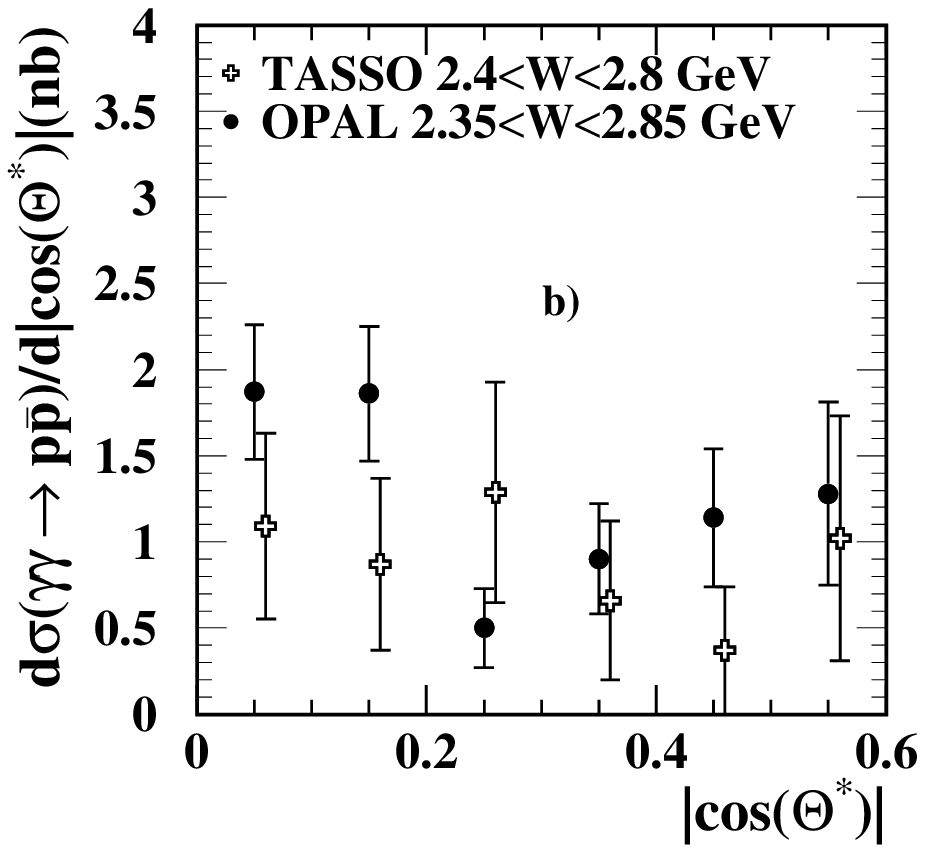}}
   \resizebox{0.35\textwidth}{!}{%
   \includegraphics{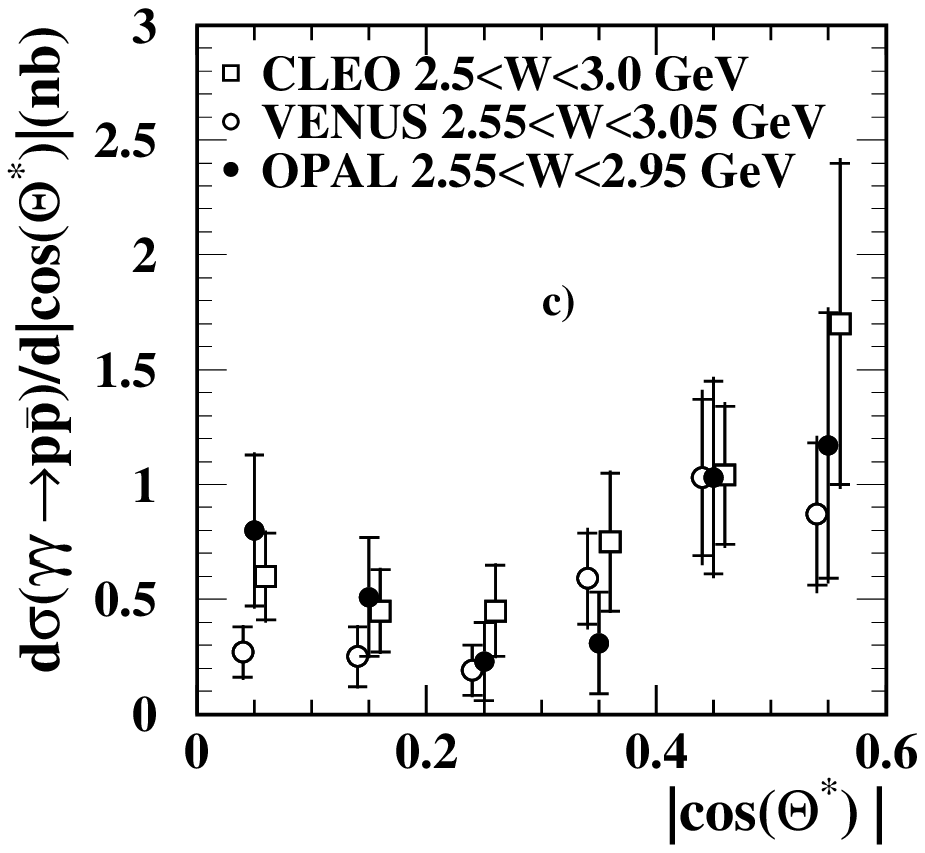}}
   \vspace*{-1.0cm}
   \caption{Differential cross-sections for $\GG\to\PP$ as 
	     a function of $\costs$ in different ranges of $W$; 
	     a, c) compared with CLEO~\cite{Artuso:1994xk}
	     and VENUS~\cite{Hamasaki:1997cy} data with statistical 
	     (inner error bars) and systematic errors (outer bars)
 and
	     b) compared with TASSO~\cite{Althoff:1983pf}.
	     The TASSO error bars are statistical only. The data points
	     are slightly displaced for clarity.} 
  \label{fig:cos1}\vspace*{-0.5cm}
 \end{figure}
The measured differential cross-sections 
${{\rm d}\SI{(\GG\to\PP)}}/{{\rm d}\costs}$ in different $W$
ranges and for $\costs<0.6$ 
are showed in Fig.~\ref{fig:cos1}.
The differential cross-section in the range $2.15<W<2.55\,\GV$  
lies below the results reported by 
VENUS~\cite{Hamasaki:1997cy} and CLEO~\cite{Artuso:1994xk} 
(Fig.~\ref{fig:cos1}a). 
Since the CLEO measurements are given for the lower $W$ range $2.0<W<2.5\,\GV$,
we rescale their results by a factor 0.635 which is the ratio of the two CLEO
total cross-section measurements integrated over the $W$ ranges
$2.0<W<2.5\,\GV$ and $2.15<W<2.55\,\GV$. This leads 
to a better agreement between the two measurements but the OPAL results
are still consistently lower.
The shapes of the $\costs$ dependence of all measurements are 
consistent apart from the highest $\costs$ bin, 
where the OPAL measurement is significantly lower than the measurements of
the other two experiments.

\begin{figure}[htbp]\vspace*{-0.5cm}
 \centering
  \resizebox{0.35\textwidth}{!}{%
  \includegraphics{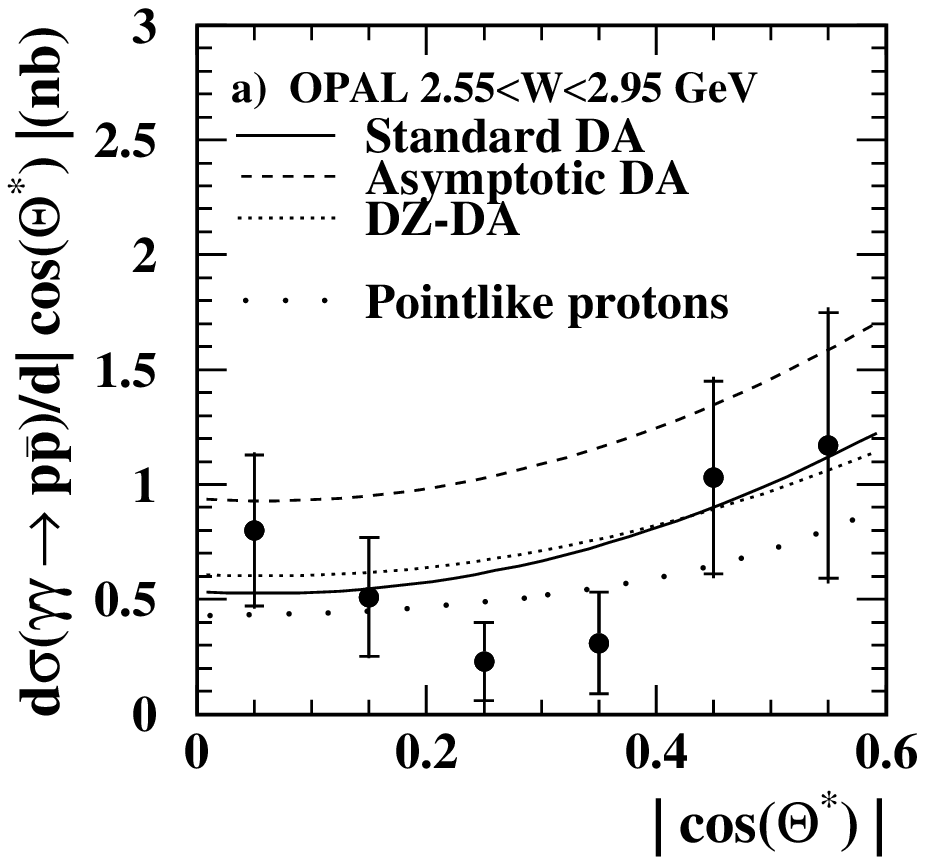}}
  \resizebox{0.35\textwidth}{!}{%
  \includegraphics{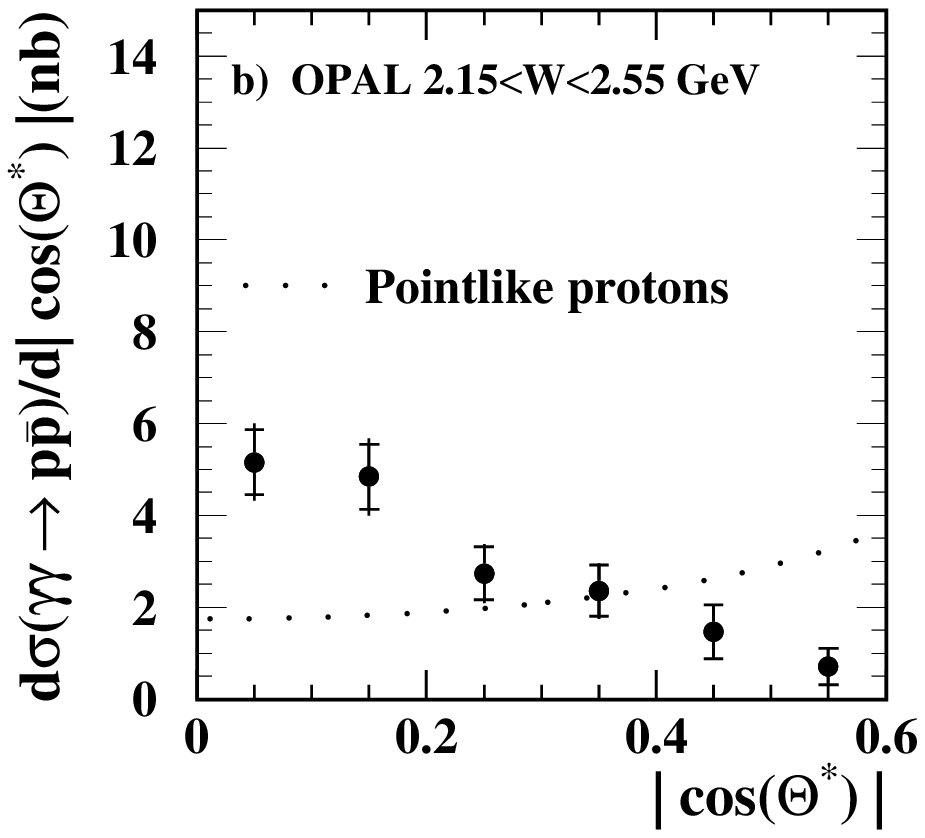}}
    \vspace*{-1.0cm}
    \caption{Measured differential cross-section, 
      ${\rm d}\SI{(\GG\to\PP)}/{\rm d}\costs$, with statistical
      (inner bars) and total uncertainties (outer bars) for 
      a) $2.55<W<2.95\,\GV$ and 
      b) $2.15<W<2.55\,\GV$. 
      The data are compared with
      the point-like approximation for the proton  
      (\ref{eq:costest}) scaled to fit the data.
      The other curves show the 
      pure quark model~\protect\cite{Farrar:1985gv},
      the diquark model of\protect~\cite{Ansel:1987vk} with
      the Dziembowski distribution amplitudes (DZ-DA), and
      the diquark model of~\cite{berger:1997} using standard
      and asymptotic distribution amplitudes.}
    \label{fig:cos2}
\vspace*{-0.5cm}
\end{figure}

In Fig.~\ref{fig:cos1}b-c the differential cross-sections 
${{\rm d}\SI{(\GG\to\PP)}}/{{\rm d}\costs}$
in the $W$ ranges $2.35<W<2.85\,\GV$ and $2.55<W<2.95\,\GV$
are compared to the 
measurements by TASSO, VENUS and CLEO in similar $W$ ranges.
The measurements are consistent within the uncertainties. 
The comparison of the differential cross-section as a function 
of $\costs$ for $2.55<W<2.95\,\GV$ with the calculation 
of~\cite{berger:1997} at $W = 2.8\,\GV$ for different distribution 
amplitudes (DA) is shown in Fig.~\ref{fig:cos2}a.
The shapes of the curves of the pure quark 
model~\cite{Farrar:1985gv,Millers:1986ca} 
and the quark-diquark model predictions~\cite{berger:1997} are consistent with 
those of the data. 

In Fig.~\ref{fig:cos2}b the differential cross-section 
${\rm d}\SI{(\GG\to\PP)}/{\rm d}\costs$ is shown versus $\costs$ 
for $2.15<W<2.55\,\GV$. 
The cross-section decreases at large $\costs$; the shape of the angular
distribution is different from that at higher $W$ values.
This indicates that for low $W$ the perturbative calculations
of~\cite{Farrar:1985gv,Millers:1986ca} are not valid.

Another important consequence of the hard scattering picture
is the hadron helicity
conservation rule. For each exclusive reaction like
$\GG\to\PP$ the sum of the two initial helicities equals
the sum of the two final ones~\cite{Brodsky:1981kj}.
According to the simplification used in~\cite{Ansel:1987vk}, 
only scalar diquarks are considered, and the (anti) proton carries the 
helicity of the single (anti) quark. 
Neglecting quark masses, quark and antiquark and hence proton and 
antiproton have to be in opposite helicity states. 
If the (anti) proton is considered as a point-like particle, simple 
QED rules determine the angular dependence of the unpolarized 
$\GG\to\PP$ differential cross-section~\cite{Budnev:1974de}: 
\begin{equation}
  \frac{{\rm d}\SI{(\GG\to\PP)}}{{\rm d}\costs} \propto \frac{(1 + \cos^{2}\theta^{*})}{(1 - \cos^{2}\theta^{*})}.
  \label{eq:costest}
\end{equation}
This expression is compared to the data in two $W$ ranges, $2.55<W<2.95\,\GV$ 
(Fig.~\ref{fig:cos2}a) and $2.15<W<2.55\,\GV$ (Fig.~\ref{fig:cos2}b). 
The normalisation in each case is determined by the best fit to 
the data. In the higher $W$ range, the prediction (\ref{eq:costest}) 
is in agreement with 
the data within the experimental uncertainties. 
In the lower $W$ range this 
simple model does not describe the data. At low $W$ soft 
processes such as meson exchange are expected to introduce other partial
waves, so that the approximations leading to (\ref{eq:costest})
become invalid~\cite{Brodsky:1987nt}.

\section{CONCLUSIONS
}
The cross-section for the process $\EE\to\EE\PP$ has been measured  
in the $\PP$ centre-of-mass energy
range of $2.15 <W< 3.95\,\GV$ using data
taken with the OPAL detector at $\SQS = 183$ and $189\,\GV$.
The measurement extends to slightly larger values of $W$
than in previous measurements.

The total cross-section $\SI(\GG\to\PP)$ as a function of $W$  
is obtained from the differential cross-section
${\rm d}\SI(\EE\to\EE\PP)/{\rm d}W$ using a luminosity
function.
For the high $\PP$ centre-of-mass energies, $W>2.3\,\GV$, the measured 
cross-section is in good agreement 
with other experimental results~\cite{Althoff:1983pf,Aihara:1987ha,Albrecht:1989hz,Artuso:1994xk,Hamasaki:1997cy}.
At lower $W$ the OPAL measurements lie below the results
obtained by CLEO~\cite{Artuso:1994xk}, and VENUS~\cite{Hamasaki:1997cy},
but agree with the JADE~\cite{Bartel:1986sy} and 
ARGUS~\cite{Albrecht:1989hz} measurements.
The cross-section  as a function of $W$ is in agreement with
the quark-diquark model predictions of~\cite{Ansel:1987vk,berger:1997}.

The power law fit yields an exponent
$n=7.5\pm0.8$ where the uncertainty is statistical only. 
Within this uncertainty,
the measurement is not able to distinguish between predictions for 
the proton to interact as a state of three quasi-free quarks 
or as a quark-diquark system.
These predictions are based on dimensional counting 
rules~\cite{Brodsky:1973kr,Matveev:1973ra}. 
 
The shape of the differential cross-section 
${\rm d}\SI{(\GG\to\PP)}/{\rm d}\costs$
agrees with the results of previous experiments in comparable
$W$ ranges, apart from the highest $\costs$ bin measured in
the range $2.15<W<2.55\,\GV$. 
In this low $W$ region 
contributions from soft processes such as meson exchange are expected
to complicate the picture by introducing extra partial waves, and
the shape of the measured differential cross-section 
${\rm d}\SI{(\GG\to\PP)}/{\rm d}\costs$
does not agree with the simple model that leads to
the helicity conservation rule.
In the high $W$ region, $2.55<W<2.95\,\GV$, the 
experimental and theoretical differential cross-sections
${\rm d}\SI{(\GG\to\PP)}/{\rm d}\costs$
agree, indicating that the data are consistent with the
helicity conservation rule.


\begin{thebibliography}{9}

\bibitem{Lepage:1980fj} G.P.\,\,Lepage\,\,\,and\,\,\,S.J.\,\,Brodsky,\,\,\,Phys.\,\,Rev. D22 (1980) 2157.
\bibitem{Farrar:1985gv} G.R. Farrar, E.~Maina and F.~Neri, Nucl.~Phys.~B259 (1985) 702.
\bibitem{Millers:1986ca} D.~Millers and J.F.~Gunion, Phys.~Rev.~D34 (1986) 2657.
\bibitem{Chernyak:1984bm} V.L.\,\,Chernyak\,\,\,and\,\,\,I.R.\,\,Zhitnitsky,\,\,\,Nucl. Phys.~B246 (1984) 52.
\bibitem{Althoff:1983pf} TASSO Collaboration, M.~Althoff et~al., Phys.~Lett.~B130 (1983) 449.
\bibitem{Bartel:1986sy} JADE Collaboration, W.~Bartel et~al., Phys.~Lett.~B174 (1986) 350.
\bibitem{Aihara:1987ha} TPC/Two Gamma Collaboration, H.~Aihara et~al., Phys.~Rev.~D36 (1987) 3506.
\bibitem{Albrecht:1989hz} ARGUS Collaboration, H.~Albrecht et~al., Z.~Phys.~C42 (1989) 543.
\bibitem{Artuso:1994xk} CLEO Collaboration, M.~Artuso et~al., Phys.~Rev.~D50 (1994) 5484.
\bibitem{Hamasaki:1997cy} VENUS Collaboration, H.~Hamasaki et~al., Phys.~Lett.~B407 (1997) 185.
\bibitem{OPALpap} OPAL Collaboration, G.~Abbiendi et al., Eur.~Phys.~J.~C28 (2003) 45.  
\bibitem{Ansel:1987vk} M.~Anselmino, P.~Kroll and B.~Pire, Z.~Phys.~C36 (1987) 89.
\bibitem{berger:1997} C.F.~Berger, B.~Lechner and W.~Schweiger, Fizika B8 (1999) 371. 
\bibitem{Anselmino:1989gu} M.~Anselmino, F.~Caruso, P.~Kroll and W.~Schweiger, Int.~J.~Mod.~Phys.~A4 (1989) 5213.
\bibitem{Kroll:1991a} P.~Kroll, M.~Sch\"{u}rmann and W.~Schweiger, Int.~J.~Mod.~Phys. A6 (1991) 4107.
\bibitem{Kroll:1993zx} P.~Kroll, Th.~Pilsner, M.~Sch\"{u}rmann and W.~Schweiger, Phys.~Lett.~B316 (1993) 546.
\bibitem{Kroll:1996pv} P.~Kroll, M.~Sch\"{u}rmann and P.A.M. Guichon, Nucl.~Phys.~A598 (1996) 435.
\bibitem{Ahmet:1991eg}
OPAL Collaboration, 
K.~Ahmet et~al., Nucl. Instr. Meth. A305 (1991) 275;
\bibitem{Akers:1995vh} R.~Akers et~al., Z. Phys. C65 (1995) 47.
\bibitem{Low:1960wv} F.E. Low, Phys. Rev. 120 (1960) 582.
\bibitem{Schuler:1996gt} G.A. Schuler,  Comp. Phys. Comm. 108 (1998) 279.
\bibitem{Lafferty:1995} G.D.~Lafferty, T.R.~Wyatt, Nucl. Instr. Meth. A355 (1995) 541.
\bibitem{Brodsky:1973kr} S.J. Brodsky and G.R. Farrar, Phys. Rev. Lett. 31 (1973) 1153.
\bibitem{Matveev:1973ra} V.A. Matveev, R.M. Muradian and A.N. Tavkhelidze, Nuovo Cim. Lett. 7 (1973) 719.
\bibitem{Brodsky:1981kj} S.J.~Brodsky and G.P.~Lepage, Phys. Rev. D24 (1981) 2848.
\bibitem{Budnev:1974de} V.M. Budnev, I.F. Ginzburg, G.V. Meledin and V.G. Serbo, Phys. Rep. 15 (1974) 181.
\bibitem{Brodsky:1987nt} S.J.~Brodsky, F.C.~Ern\'{e}, P.H.~Damgaard and P.M.~Zerwas, Contribution to ECFA Workshop LEP200, Aachen, Germany, Sep 29 - Oct 1, 1986.
\end{thebibliography}
\end{document}